
\def\pmb#1{\setbox0=\hbox{#1}%
  \hbox{\kern-.025em\copy0\kern-\wd0
  \kern.05em\copy0\kern-\wd0
  \kern-0.025em\raise.0433em\box0} }

\catcode`@=11
\def\leftrightarrowfill{$\m@th\mathord\leftarrow \mkern-6mu
  \cleaders\hbox{$\mkern-2mu \mathord- \mkern-2mu$}\hfill
  \mkern-6mu \mathord\rightarrow$}
\def\overleftrightarrow#1{\vbox{\ialign{##\crcr
     \leftrightarrowfill\crcr\noalign{\kern-1pt\nointerlineskip}
     $\hfil\displaystyle{#1}\hfil$\crcr}}}
\catcode`@=12

\def\approxge{\hbox {\hfil\raise .4ex\hbox{$>$}\kern-.75 em
\lower .7ex\hbox{$\sim$}\hfil}}
\def\approxle{\hbox {\hfil\raise .4ex\hbox{$<$}\kern-.75 em
\lower .7ex\hbox{$\sim$}\hfil}}
\def \abstract#1 {\vskip 0.5truecm\sepline\vskip 0.5truecm
$$\vbox{\hsize=15truecm\noindent #1}$$}
\def \SISSA#1#2 {\vfil\vfil\centerline{Ref. S.I.S.S.A. #1 CM (#2)}}
\def \PACS#1 {\vfil\line{\hfil\hbox to 15truecm{PACS numbers: #1 \hfil}\hfil}}
\def \hfigure
     #1#2#3       {\midinsert \vskip #2 truecm $$\vbox{\hsize=14.5truecm
             \seven\baselineskip=10pt\noindent {\bcp \noindent Figure  #1}.
                   #3 } $$ \vskip -20pt \endinsert }
\def \hfiglin
     #1#2#3       {\midinsert \vskip #2 truecm $$\vbox{\hsize=14.5truecm
              \seven\baselineskip=10pt\noindent {\bcp \hfil\noindent
                   Figure  #1}. #3 \hfil} $$ \vskip -20pt \endinsert }
\def \vfigure
     #1#2#3#4     {\dimen0=\hsize \advance\dimen0 by -#3truecm
                   \midinsert \vbox to #2truecm{ \seven
                   \parshape=1 #3truecm \dimen0 \baselineskip=10pt \vfill
                   \noindent{\bcp Figure #1} \pretolerance=6500#4 \vfill }
                   \endinsert }

\def \ref
     #1#2         {\smallskip \item{[#1]}#2}
\def \sepline     {\medskip\centerline{\vbox{\hrule width5truecm}} \medskip}

\def \tabrul2     {\noalign{\vskip 5truept \hrule \vskip 2truept \hrule
                   \vskip 5truept} }
\footline={\ifnum\pageno>0 \tenrm \hss \folio \hss \fi }
\def\today
 {\count10=\year\advance\count10 by -1900 \number\day--\ifcase
  \month \or Jan\or Feb\or Mar\or Apr\or May\or Jun\or
             Jul\or Aug\or Sep\or Oct\or Nov\or Dec\fi--\number\count10}
\def\hour{\count10=\time\count11=\count10
\divide\count10 by 60 \count12=\count10
\multiply\count12 by 60 \advance\count11 by -\count12\count12=0
\number\count10 :\ifnum\count11 < 10 \number\count12\fi\number\count11}
\def\draft{
   \baselineskip=20pt
   \def\makeheadline{\vbox to 10pt{\vskip-22.5pt
   \line{\vbox to 8.5pt{}\the\headline}\vss}\nointerlineskip}
   \headline={\hfill \seven {\bcp Draft version}: today is \today\ at \hour
              \hfill}
          }
\catcode`@=11
%
%
\def\b@lank{ }
\newif\if@simboli
\newif\if@riferimenti
\newwrite\file@simboli
\def\simboli{
    \immediate\write16{ !!! Genera il file \jobname.SMB }
    \@simbolitrue\immediate\openout\file@simboli=\jobname.smb}
\newwrite\file@ausiliario
\def\riferimentifuturi{
    \immediate\write16{ !!! Genera il file \jobname.AUX }
    \@riferimentitrue\openin1 \jobname.aux
    \ifeof1\relax\else\closein1\relax\input\jobname.aux\fi
    \immediate\openout\file@ausiliario=\jobname.aux}
\newcount\eq@num\global\eq@num=0
\newcount\sect@num\global\sect@num=0
\newif\if@ndoppia
\def\numerazionedoppia{\@ndoppiatrue\gdef\la@sezionecorrente{\the\sect@num}}
\def\se@indefinito#1{\expandafter\ifx\csname#1\endcsname\relax}
\def\spo@glia#1>{} 
\newif\if@primasezione
\@primasezionetrue
\def\s@ection#1\par{\immediate
    \write16{#1}\if@primasezione\global\@primasezionefalse\else\goodbreak
    \vskip\spaziosoprasez\fi\noindent
    {\bf#1}\nobreak\vskip\spaziosottosez\nobreak\noindent}
%
\def\sezpreset#1{\global\sect@num=#1
    \immediate\write16{ !!! sez-preset = #1 }   }
\def\spaziosoprasez{50pt plus 60pt}
\def\spaziosottosez{15pt}
\def\sref#1{\se@indefinito{@s@#1}\immediate\write16{ ??? \string\sref{#1}
    non definita !!!}
    \expandafter\xdef\csname @s@#1\endcsname{??}\fi\csname @s@#1\endcsname}
\def\adv#1{\global\advance\sect@num by #1
           \global\eq@num=0}
\def\autosez#1#2\par{
    \global\advance\sect@num by 1\if@ndoppia\global\eq@num=0\fi
    \xdef\la@sezionecorrente{\the\sect@num}
    \def\usa@getta{1}\se@indefinito{@s@#1}\def\usa@getta{2}\fi
    \expandafter\ifx\csname @s@#1\endcsname\la@sezionecorrente\def
    \usa@getta{2}\fi
    \ifodd\usa@getta\immediate\write16
      { ??? possibili riferimenti errati a \string\sref{#1} !!!}\fi
    \expandafter\xdef\csname @s@#1\endcsname{\la@sezionecorrente}
    \immediate\write16{\la@sezionecorrente. #2}
    \if@simboli
      \immediate\write\file@simboli{ }\immediate\write\file@simboli{ }
      \immediate\write\file@simboli{  Sezione
                                  \la@sezionecorrente :   sref.   #1}
      \immediate\write\file@simboli{ } \fi
    \if@riferimenti
      \immediate\write\file@ausiliario{\string\expandafter\string\edef
      \string\csname\b@lank @s@#1\string\endcsname{\la@sezionecorrente}}\fi
    \goodbreak\vskip 48pt plus 60pt
\centerline{\lltitle #2}                     
\par\nobreak\vskip 15pt \nobreak\noindent}
\def\semiautosez#1#2\par{
    \gdef\la@sezionecorrente{#1}\if@ndoppia\global\eq@num=0\fi
    \if@simboli
      \immediate\write\file@simboli{ }\immediate\write\file@simboli{ }
      \immediate\write\file@simboli{  Sezione ** : sref.
          \expandafter\spo@glia\meaning\la@sezionecorrente}
      \immediate\write\file@simboli{ }\fi
\noindent\lltitle \s@ection#2 \par}
\def\eqpreset#1{\global\eq@num=#1
     \immediate\write16{ !!! eq-preset = #1 }     }
\def\eqref#1{\se@indefinito{@eq@#1}
    \immediate\write16{ ??? \string\eqref{#1} non definita !!!}
    \expandafter\xdef\csname @eq@#1\endcsname{??}
    \fi\csname @eq@#1\endcsname}
\def\eqlabel#1{\global\advance\eq@num by 1
    \if@ndoppia\xdef\il@numero{\la@sezionecorrente.\the\eq@num}
       \else\xdef\il@numero{\the\eq@num}\fi
    \def\usa@getta{1}\se@indefinito{@eq@#1}\def\usa@getta{2}\fi
    \expandafter\ifx\csname @eq@#1\endcsname\il@numero\def\usa@getta{2}\fi
    \ifodd\usa@getta\immediate\write16
       { ??? possibili riferimenti errati a \string\eqref{#1} !!!}\fi
    \expandafter\xdef\csname @eq@#1\endcsname{\il@numero}
    \if@ndoppia
       \def\usa@getta{\expandafter\spo@glia\meaning
       \la@sezionecorrente.\the\eq@num}
       \else\def\usa@getta{\the\eq@num}\fi
    \if@simboli
       \immediate\write\file@simboli{  Equazione
            \usa@getta :  eqref.   #1}\fi
    \if@riferimenti
       \immediate\write\file@ausiliario{\string\expandafter\string\edef
       \string\csname\b@lank @eq@#1\string\endcsname{\usa@getta}}\fi}
\def\autoeqno#1{\eqlabel{#1}\eqno(\csname @eq@#1\endcsname)}
\def\autoleqno#1{\eqlabel{#1}\leqno(\csname @eq@#1\endcsname)}
\def\eqrefp#1{(\eqref{#1})}
\def\eq{\autoeqno}
\def\req{\eqrefp}
\newcount\cit@num\global\cit@num=0
\newwrite\file@bibliografia
\newif\if@bibliografia
\@bibliografiafalse
\def\lp@cite{[}
\def\rp@cite{]}
\def\trap@cite#1{\lp@cite #1\rp@cite}
\def\lp@bibl{[}
\def\rp@bibl{]}
\def\trap@bibl#1{\lp@bibl #1\rp@bibl}
\def\refe@renza#1{\if@bibliografia\immediate        
    \write\file@bibliografia{
    \string\item{\trap@bibl{\cref{#1}}}\string
    \bibl@ref{#1}\string\bibl@skip}\fi}
\def\ref@ridefinita#1{\if@bibliografia\immediate\write\file@bibliografia{
    \string\item{?? \trap@bibl{\cref{#1}}} ??? tentativo di ridefinire la
      citazione #1 !!! \string\bibl@skip}\fi}
\def\bibl@ref#1{\se@indefinito{@ref@#1}\immediate
    \write16{ ??? biblitem #1 indefinito !!!}\expandafter\xdef
    \csname @ref@#1\endcsname{ ??}\fi\csname @ref@#1\endcsname}
\def\c@label#1{\global\advance\cit@num by 1\xdef            
   \la@citazione{\the\cit@num}\expandafter
   \xdef\csname @c@#1\endcsname{\la@citazione}}
\def\bibl@skip{\vskip +4truept}
\def\stileincite#1#2{\global\def\lp@cite{#1}\global   
    \def\rp@cite{#2}}                                 
\def\stileinbibl#1#2{\global\def\lp@bibl{#1}\global   
    \def\rp@bibl{#2}}                                 
\def\citpreset#1{\global\cit@num=#1
    \immediate\write16{ !!! cit-preset = #1 }    }
\def\autobibliografia{\global\@bibliografiatrue\immediate
    \write16{ !!! Genera il file \jobname.BIB}\immediate
    \openout\file@bibliografia=\jobname.bib}
\def\cref#1{\se@indefinito                  
   {@c@#1}\c@label{#1}\refe@renza{#1}\fi\csname @c@#1\endcsname}
\def\cite#1{\trap@cite{\cref{#1}}}                  
\def\ccite#1#2{\trap@cite{\cref{#1},\cref{#2}}}     
\def\ncite#1#2{\trap@cite{\cref{#1}--\cref{#2}}}    
\def\upcite#1{$^{\,\trap@cite{\cref{#1}}}$}               
\def\upccite#1#2{$^{\,\trap@cite{\cref{#1},\cref{#2}}}$}  
\def\upncite#1#2{$^{\,\trap@cite{\cref{#1}-\cref{#2}}}$}  

\def\clabel#1{\se@indefinito{@c@#1}\c@label           
    {#1}\refe@renza{#1}\else\c@label{#1}\ref@ridefinita{#1}\fi}
\def\biblskip#1{\def\bibl@skip{\vskip #1}}           
\def\insertbibliografia{\if@bibliografia             
    \immediate\write\file@bibliografia{ }
    \immediate\closeout\file@bibliografia
    \catcode`@=11\input\jobname.bib\catcode`@=12\fi}
\def\commento#1{\relax}
\def\biblitem#1#2\par{\expandafter\xdef\csname @ref@#1\endcsname{#2}}
\catcode`@=12
\tolerance 100000
\biblskip{+8truept}                        
\def\hbup{\hfill\break\baselineskip 14pt}  
\global\newcount\notenumber \global\notenumber=0
\def\note #1 {\global\advance\notenumber by1 \baselineskip 10pt
              \footnote{$^{\the\notenumber}$}{\nine #1} \interlinea}


 
\def\gtrsim{\ \rlap{\raise 2pt \hbox{$>$}}{\lower 2pt \hbox{$\sim$}}\ }
\def\lesssim{\ \rlap{\raise 2pt \hbox{$<$}}{\lower 2pt \hbox{$\sim$}}\ }

\def\bs{\bigskip}

\def\no{\noindent}

\def\ea{{\elevenit et.al.}}

\def\npb#1{{\elevenit Nucl. Phys.} {\elevenbf B#1},}
\def\plb#1{{\elevenit Phys. Lett.} {\elevenbf B#1},}
\def\prd#1{{\elevenit Phys. Rev.} {\elevenbf D#1},}
\def\prl#1{{\elevenit Phys. Rev. Lett.} {\elevenbf #1},}

\def\prep#1{{\elevenit Phys. Rep.} {\elevenbf #1},}

\stileincite{}{}     
\stileinbibl{}{.}    





\global\newcount\notenumber \global\notenumber=0
\def\note #1 {\global\advance\notenumber by1 \baselineskip 12pt
              \footnote{$^{\the\notenumber}$}{\nine #1} \interlinea}

\def\gtrsim{\ \rlap{\raise 2pt \hbox{$>$}}{\lower 2pt \hbox{$\sim$}}\ }
\def\lesssim{\ \rlap{\raise 2pt \hbox{$<$}}{\lower 2pt \hbox{$\sim$}}\ }
\def\npb{Nucl. Phys. B }
\def\plb{Phys. Lett. B }
\def\prd{Phys. Rev. D }
\def\prl{Phys. Rev. Lett. }

\def\prep{Phys. Rep. }

\autobibliografia    
\def\bs{\bigskip}

\def\t{\tilde}

\font\tenrm=cmr10
\font\tenit=cmti10
\font\elevenbf=cmbx10 scaled\magstep 1
\font\elevenrm=cmr10 scaled\magstep 1
\font\elevenit=cmti10 scaled\magstep 1


\line{\hfil }
\vglue 1cm
\hsize=6.0in
\vsize=8.5truein
\parindent=3pc
\baselineskip=10pt
\rightline{FTUV/94--1}
\medskip
\bs\bs\bs\bs
\centerline{\elevenbf DARK MATTER AND SOLAR NEUTRINOS IN SUSY$^\dagger$}
\bs\bs
\def\authors{
                 \centerline{\tenrm DANIELE TOMMASINI}
\smallskip
\centerline{\tenit Departament de F\'\i sica Te\` orica, Universitat de
                                                      Val\` encia}
\centerline{\tenit Instituto de F\'\i sica Corpuscular - C.S.I.C.}
\centerline{\tenit 46100 Burjassot, Val\` encia, SPAIN}
}
\authors
\bs\bs
\medskip
\centerline{\tenrm ABSTRACT}
\bs
 \leftskip=3pc
 \tenrm\baselineskip=12pt
In the Supersymmetric extension of the
Standard Model with minimal particle
content the three neutrinos can have non trivial masses and mixings,
generated at 1 loop due to renormalizable lepton number violating
interactions.  We show that the resulting mass matrix can provide
simultaneously a significant amount of
the Dark Matter of the Universe and solve the solar neutrino problem, if
the free parameters of the model are fixed to values which are consistent
with all the present accelerator and cosmological constraints. The theory
also predicts new effects in future experiments looking
for neutrino oscillations.
 \vfill\noindent
--------------------------------------------\phantom{-} \hfil\break
\medskip
\leftline{$^\dagger$ Talk given at the
{\elevenit ``International School on Cosmological Dark Matter''},}
\leftline{\phantom{$^\dagger$} Valencia, Spain, October 4-8, 1993.}
\vglue 1truecm
\leftline{E-mail: 16444::tommasin, tommasin@evalvx}
\bigskip
\centerline{January 1994}
\par\vfill\eject
\vglue 0.8cm
\def\interlinea{\baselineskip=14pt}
\baselineskip=14pt
\elevenrm

Cosmological and astrophysical considerations strongly suggest
that neutrinos are massive particles, in spite of the present
lack of any accelerator evidence.

First, neutrinos are the only known particles which can
contribute significantly to the Dark Matter (DM) of the Universe if the
sum of their masses is in the $\sim10$ eV range. Recent indications
\upcite{hotdm}
suggest that hot DM, such as neutrino DM, should compose the $\sim30$ \%
of the Universe, corresponding to $\sum_{\nu}m_{\nu}\simeq7$ eV.

Second, the experimental measurements of the
flux of the neutrinos from the sun imply \upcite{bludman93}
non trivial mass matrices
\upcite{msw} and/or nonstandard interactions \upcite{mswfcnc}
for the neutrinos.
The simplest and most appealing solution to this ``solar neutrino
problem" is provided by the Mikheyev--Smirnov--Wolfenstein (MSW) \upcite{msw}
scenario, which describes all the observations assuming that
the electron neutrino $\nu_e$ is resonantly converted into
another state such as $\nu_\mu$ or $\nu_\tau$, resulting in a reduction of
the observed flux in the earth. The allowed region for the squared mass
difference and mixing angle between the two states is
$$
\Delta m^2 =(0.5-1.2)\times10^{-5}{\rm eV}^2, \qquad
\sin^2 2\theta=(0.3-1.0)\times10^{-2},
\eq{dmsn1}
$$
in the case of small vacuum mixing \upcite{smirnov93}.

In the Standard Model (SM) of Particle Physics, neutrinos are strictly
massless as a consequence of the gauge symmetry and particle content of the
theory. In extensions of the SM, neutrinos can acquire a mass due to the
presence of sterile neutrinos, and in this case the smallness of the
mass of the known neutrino states can be elegantly explained by the
see-saw mechanism \upcite{seesaw}. The resulting mass matrix among the
light states can then account both for the hot Dark Matter (e.g.
with $m_{\nu_\tau}\sim10$ eV) and for the MSW solution of
the solar neutrino problem (e.g. by $\nu_e-\nu_\mu$ oscillations)
\upcite{bludman}.

In some very popular extensions of the SM, such as the
Supersymmetric SM (SSM) with minimal particle content and the
simplest SUSY-SU(5) theory, no nonstandard neutral fermion is present,
so that the see-saw mechanism cannot be implemented. However, in the
supersymmetric case the SU(2)$\times$U(1) gauge symmetry and the
minimal particle content allow for renormalizable lepton number violating
operators \upcite{susynmm}, in contrast to the non supersymmetric case.
The three left handed neutrino flavours acquire then a non trivial
mass matrix, and the heaviest state can be naturally in the $\sim10$ eV
range \upcite{cnd}. We will study here the possibility that this happens
while the two lighter
components have masses and mixing such as to account also for the MSW
mechanism, in the framework of the SSM with minimal particle content,
making just a minimum number of naturalness assumptions in order to
minimize the number of free parameters. The result is that
it is possible to solve simultaneously the DM and solar neutrino problems
just by supersymmetrizing the SM and allowing for the most general
superpotential, which involve lepton number violating interactions,
without contradicting any accelerator or cosmological constraint,
and this solution imply predictions which will be tested in
the future experiments looking for neutrino
oscillations.

\bigskip

In the Supersymmetric Standard Model (SSM) \upcite{ssm} with minimal particle
content, lepton number violation can arise in the superpotential through
renormalizable terms \upcite{susynmm},
involving either two lepton and an antilepton
or quark--antiquark--lepton, chiral (``left--handed") superfields.
Fig. 1 shows the generic one loop diagrams
that contribute to the mass entry $m_{\nu_i\nu_j}$
($i,j,k=1,...,3$ are the flavour indices).
The fermionic $f_{k'}$ and scalar $\tilde f_k\tilde f^c_k$
internal lines can be either charged $\pm1$
lepton--slepton pairs,
or charged $\pm1/3$ quark--squark pairs.
Diagrams (a) and (b) must be considered together, and summed, because they are
proportional to the same product of lepton number violating coupling constants;
for the diagonal
matrix elements ($i=j$), if the two flavors running in the loop are
the same ($k=k'$), the two diagrams coincide, so that just
one should be considered.

In the diagram of fig. 1 an helicity flip
on the internal fermion line is necessary. As explicitly indicated, this
also requires a mixing of the scalars $\t f_k$ and $\t f_k^c$,
described by the ``mass insertions" $\Delta_k$
\upcite{susynmm}.
For instance, in the usual ``Minimal" SSM (MSSM), with soft supersymmetry
breaking arising from low energy supergravity, every
contribution to the mixing $\Delta_k$ is proportional to
the mass $m_k$ of the fermionic superpartner $f_k$ of $\t f_k$,
$$\Delta_{k}=\t m m_k,
\eq{mssm}
$$
where $\t m$ is a typical supersymmetry--breaking mass parameter
\upcite{susynmm}.
For simplicity, we will assume in the following that eq. \req{mssm} holds.

The generic single diagram of fig. 1 contributes to  $m_{\nu_i\nu_j}$
as
$$\delta m_{\nu_i\nu_{j}}\simeq
N_c{\lambda_{jk{k'}}\lambda_{i{k'}k}\over 16\pi^2}m_{k'}
{\Delta_{k} \over \t m_k^2},
\eq{mij}
$$
where $N_c$ is a colour factor ($N_c=3$ for the diagrams with quark
flavors, $N_c=1$ otherwise),
$m_{k'}$ is the mass of the internal charged fermion $f_{k'}$,
and $\t m_k^2$ is the average mass of the two mass eigenstates resulting from
the mixing between $\t f_k$ and $\t f^c_k$.
Of course $m_{k'}/\t m_k\ll1$ and we are also making an expansion in
$\Delta_{k} /\t m_k^2$.
The lepton number breaking coupling constants $\lambda_{ihk}$
for the purely leptonic vertices, $\lambda_{ihk}\equiv\lambda_{ihk}^l$, are
antisymmetric in the indices $h,k$, and they differ in general from the
couplings for the lepton--quark--squark vertices,
$\lambda_{ihk}\equiv\lambda_{ihk}^q$. The sets of the $\lambda^l$
and $\lambda^q$ consist then of 18 and 27 free parameters, respectively.
We stress again that there is no compelling {\elevenit a priori} reason to put
all
these coupling constants to zero, assuming the conservation of
a symmetry generalizing
lepton number in the supersymmetric case, called R-parity,
since this is no longer a result of the gauge symmetry, unlike in the
non supersymmetric SM.
We will make the reasonable assumption that any hierarchy amongst the coupling
constants $\lambda_{ihk}$ can be justified by an (approximate)
flavour symmetry.
In particular, we assume that all the coupling constants $\lambda$
that violate only a single lepton flavour number, say $L_i=L_e$,
are of the
same order, their value measuring the amount of breaking of the symmetry
$L_i$. The remaining coupling constants, $\lambda^l_{123}$, $\lambda^l_{132}$
and $\lambda^l_{231}$,
arise when all the lepton numbers are broken, so that they are expected to
be not larger than any of the others.
This implies that for any mass entry $m_{\nu_i\nu_j}$ there is just
one dominating contribution from fig. 1, corresponding to the
largest product $m_{k'}\Delta_k=m_{k'}\t m m_k$, that is to the exchange of
bottom quarks and squarks. The full $3\times3$ neutrino mass matrix depends
then just
on three parameters, $\lambda_e=\lambda_{133}^q$, $\lambda_\mu=\lambda_{233}^q$
and $\lambda_\tau=\lambda_{333}^q$, as
$$
{\bf m}={N_c\over 16\pi^2}{m_b^2\t m \over \t m_k^2}
\pmatrix{
{\lambda_e}^2 & 2\lambda_e\lambda_\mu & 2\lambda_e\lambda_\tau\cr
2\lambda_e\lambda_\mu & {\lambda_\mu}^2 & 2\lambda_\mu\lambda_\tau\cr
2\lambda_e\lambda_\tau & 2\lambda_\mu\lambda_\tau & {\lambda_\tau}^2\cr
}.\eq{mijq}
$$

Let us assume that one neutrino state constitutes the hot DM with
a mass $\sim7$ eV, and that the remaining two states have
a mass difference and mixing angle given by eq. \req{dmsn1} corresponding
to the MSW solution of the solar neutrino problem.
These properties should result from the diagonalization of the mass matrix
$m$ in eq. \req{mijq}, so that all the three independent parameters
$\lambda_e{\t m/\t m_b^2}$, $\lambda_\mu{\t m/\t m_b^2}$ and
$\lambda_\tau{\t m/\t m_b^2}$ will be determined. The non
trivial fact is that the solution will correspond to {\elevenit experimentally
allowed} values for the $\lambda$s. Furthermore, once the parameters
$\lambda$s are
fixed, all the entries of the mass matrix \req{mijq} will be determined,
and we will have unambiguous predictions for the mixing angles in the two
channels that do not correspond to the MSW mixing.

Let us consider the case $\lambda_\tau>\lambda_\mu>\lambda_e$, that is
$L_\tau$ is violated first, providing a DM neutrino $\nu_{\rm
DM}\simeq\nu_\tau$ of mass $m_{\nu_{\rm DM}}\simeq7$ eV. Then the breaking
of $L_\mu$ provides the MSW mass scale, and finally the smaller violation
of $L_e$ generates also the MSW mixing of eq. \req{dmsn1}.
{}From the diagonalization of the matrix $m$ in eq. \req{mijq} we get then
$$
\lambda_\tau\sqrt{\t m\over100 {\rm GeV}}\left({\rm 100GeV}\over \t m_b
\right)\simeq 10^{-3},
$$
$$
\lambda_\mu\sqrt{\t m\over100 {\rm GeV}}\left({\rm 100GeV}\over \t m_b
\right)\simeq 10^{-5},
\eq{dmsn5}
$$
$$
\lambda_e\sqrt{\t m\over100 {\rm GeV}}\left({\rm 100GeV}\over
\t m_b\right)\simeq 5\times10^{-7}.
$$
These values are consistent with the present accelerator
and astrophysical limits \upcite{rplimits}.
Furthermore, by diagonalizing the matrix \req{mijq} corresponding to
these values for the $\lambda$s one gets automatically
$\nu_\mu-\nu_\tau$ oscillations with
$\sin^2 2\theta\simeq1.3\times10^{-3}$,
3 times below the present limit \upcite{pdg92}.
The predicted $\nu_e-\nu_\tau$ mixing is negligibly small,
$\sin^2 2\theta\simeq\times10^{-6}$.

Similar conclusions hold
if for some reason the vertices $\lambda^h$ are suppressed with respect
to the purely leptonic couplings $\lambda^l$. In this case, the largest
diagrams involve $\tau-\t \tau$ exchange, and the heavy DM should be
mostly $\nu_\mu$\upcite{cnd},
since the entry $m_{\nu_\tau\nu_\tau}$ of the
mass matrix involve the exchange of muonic flavours in the loop due to the
antisymmetric character of the $\lambda^l$s.

We see that, in order to account for both the hot DM and the MSW
mechanism, a three order of
magnitude hierarchy is needed in eq. \req{dmsn5} between the $L_\tau$
violating couplings, $\lambda_\tau$, and the $L_e$ violating couplings
$\lambda_e$. This is comparable to the hierarchy between the charged lepton
masses $m_\tau$ and $m_e$.
We notice also that the values in eq. \req{dmsn5} do not satisfy the
stringent cosmological bound
$\vert\lambda\vert\lesssim2\times10^{-7}(\t m_b/100{\rm GeV})^{1/2}$,
obtained from the requirement that the primordial Baryon Asymmetry of the
Universe (BAU) be
not washed out \upcite{cosmlimlam}. However, this bound depends on the
assumption that the BAU was generated at the GUT scale, which may not be
the case. In particular, it has been shown that the BAU could have been
produced just below the electroweak scale due to the (possible)
lepton number violating interactions themselves, and the required values
\upcite{masiero-riotto}
for the $\lambda$s can agree with eq. \req{dmsn5} for some special choice
of the supersymmetric mass parameters.

We conclude noticing that in this framework it is impossible to account
also for the atmospheric neutrino data \upcite{smirnov93},
if the indication of non vanishing
mixing is confirmed. In general, it is very hard to reconcile
hot DM, MSW and atmospheric
neutrino mixing with just the three standard left handed neutrinos.
If all these phenomena were taken seriously,
this could be an hint for some more light neutral state, such as a
singlet neutrino \upcite{dmsnath}.

\bigskip

I would like to thank E. Akhmedov, F. de Campos,
A. Masiero, E. Nardi, J. Peltoniemi,
E. Roulet, G. Senjanovic and J. Valle for useful discussions.

\par\vfill\eject

\def\bi{\biblitem}
\def\hbup{\hfill\break\baselineskip 12pt}
\null
\baselineskip 8pt
\centerline{\elevenbf References}
\vskip .8truecm
\biblitem{pdg92}
Particle Data Group, \prd 45 (1992) 1. \par
\bi{dmsnath}
J. Peltoniemi, D. Tommasini and J.W.F. Valle, \plb 298 (1993) 383.\par
\bi{susy-fcnc}
F. Zwirner, \plb 132 (1983) 103;\hbup
R. Barbieri and A. Masiero, \npb 267 (1986) 679;\hbup
G.G. Ross and J.W.F. Valle, \plb 151 (1985) 375;\hbup
J. Ellis, G. Gelmini, C. Jarlskog, G.G. Ross and J.W.F. Valle, \plb 150
(1985) 142;\hbup
F. Gabbiani and A. Masiero, \npb 259 (1991) 323.\par
\bi{solarnumagmom}
M. Voloshin, M. Vysotskii and L.B. Okun, JETP 64 (1986) 446;
Sov. J. Nucl. Phys. 44 (1986) 440.\par
\bi{rpbreak}
C.S. Aulakh and R.N. Mohapatra, \prd {119} (1983) 136;
F. Zwirner, \plb  {132} (1983) 103; L.J. Hall and M. Suzuki, \npb
{231} (1984) 419; I.H. Lee, \npb  {246} (1984) 120;
G.G. Ross and J.W.F. Valle, \plb  {151}
(1985) 375; J. Ellis et al., \plb   {150}
(1985) 142; S. Dawson \npb  {261} 297; R. Barbieri and A.
Masiero, \npb  {267} (1986) 679; R.N. Mohapatra, \prd
{11} (1986) 3457;
V. Barger, G. Giudice and T.Y. Han, \prd  {40} (1989)
2987; E. Ma and P. Roy \prd  {41} (1990)  988;
E. Ma and D. Ng, \prd  {41} (1990) 1005;
S. Dimopoulos, R. Esmailzadeh, L.J. Hall, J.P. Merlo, G.D. Sparkman,
\prd {41} (1990) 2099; L.H. Hall, Mod. Phys. Lett.
A {5} (1990) 467.\par
\bi{mswfcnc}
E. Roulet, \prd 44 (1991) R935.\par
\bi{seesaw}
M. Gell-Mann, P. Ramond a d R. Slansky, in {\elevenit Supergravity}, ed.
F. van Niewenhuizen and D. Freedman, North Holland, Amsterdam (1979);\hbup
T. Yanagida, in Proc. of the Workshop on the {\elevenit Unified Theory and
Baryon
Number in the Universe}, eds. O. Sawada and A. Sugamoto (KEK, Tsukuba)
(1979);\hbup
R.N. Mohapatra and G. Senjanovic, \prl 44 (1980) 912.\par
\bi{bludman}
S. Bludman, N. Hata, D.C. Kennedy and P.G. Langacker, \prd 47 (1993)
2220.\par
\bi{bludman93}
S. Bludman, N. Hata and P.G. Langacker, UPR--0572--T (1993). \par
\bi{72}
K.S. Babu and R.N. Mohapatra, \prl  {64}
(1990) 1705. \par
\bi{79}
K Fujikawa and R. Shrock, \prl {45} (1980) 963.\par
\bi{73}
R. Barbieri and G. Fiorentini, \npb {304} (1988) 909.\par
\bi{ssm}
For a review, see H.P. Nilles, \prep {110}
(1984) 1, and references therein.\par
\bi{rplimits}
V. Barger, G. Giudice and T.Y. Han, \prd  {40} (1989)
2987;\hbup
H. Dreiner and G.G. Ross, \npb 365 (1991) 597;\hbup
K. Enqvist, A Masiero and A. Riotto, \npb 373 (1992) 95.\par
\bi{76}
M. Voloshin, Yad. Fiz.  {48} (1988) 804
[Sov. J. Nucl. Phys.  {48} (1988) 512];
R. Barbieri and R. N. Mohapatra, \plb   {218} (1989) 225;
T. Liu, \plb   {225} (1989) 148;
K.S. Babu and R.N. Mohapatra, \prl  {63}
(1989) 228; G. Ecker, W. Grimus and H. Neufeld, \plb   {232}
(1989) 217; D. Chang, W. Keung and G.
Senjanovic, \prd  42 (1990) 1599;
M.I. Vysotsky, ICTP preprint IC/90/78 (1990).\par
\bi{77}
M. Leurer and N. Marcus, \plb  {237} (1990) 81.\par
\bi{72b}
K.S. Babu and R.N. Mohapatra, \prl  {64}
(1990) 1705;
K.S. Babu and R.N. Mohapatra, \prd 42 (1990) 3778.\par
\bi{ddm}
A. De R\'ujula and S.L. Glashow, \prl {45} (1980) 942;
A.L. Mellot and D.W. Sciama, \prl {46} (1981) 1369;
Y. Rephaeli and A.S. Szalay, \plb  106 (1981) 73;
D.W. Sciama, Mon. Not. Roy. As. Soc. 198 (1982) 1;
Mon. Not. Roy. As. Soc. 200 (1982) 13; \prl  65 (1990) 2839; Nature 346
(1990) 40; {\it ibid. 348 } (1990) 617; Mon. Not. Roy. As. Soc. 244 (1990)
1; {\it ibid. 246 } (1990) 191; Astroph. J. 364 (1990) 549;
A.L. Mellot, D.W. McKay and J.P. Ralston, Astroph. J. 324 (1988) L43;
P. Salucci and D.W. Sciama, Mon. Not. Roy. As. Soc. 244 (1990) 9;
D.W. Sciama and P. Salucci, Mon. Not. Roy. As. Soc. 247 (1990) 506;
D.W. Sciama, \npb  (Proc. Suppl.) 19 (1991) 138;
S. Dodelson and J.M. Jubas, \prd  45 (1992) 1076. \par
\bi{sciama}
D.W. Sciama in ref. \par
\bi{81}
A. De R\'ujula and S.L. Glashow, \prl {45} (1980) 942.\par
\bi{82}
A.L. Mellot and D.W. Sciama, \prl {46} (1981) 1369.\par
\bi{83}
D.W. Sciama, preprint SISSA 29 A (1990).\par
\bi{84}
D.W. Sciama, preprint SISSA 28 A (1990).\par
\bi{85}
S. Tremaine and J.E. Gunn, \prl {42} (1979) 407;
D.W. Sciama, Mon. Not. Roy. As. Soc. 246 (1990) 191.\par
\bi{86}
A. Zee, \plb  {93} (1980) 389;
S. Petcov, \plb  {115} (1982) 401;
R.N. Mohapatra, \plb  {201} (1988) 517;
J.P. Ralston, D.W. McKay and A.L. Mellot, \plb  {202} (1988)
40.\par
\bi{89}
See for instance L. Wolfenstein, \plb  {107} (1981) 77;
B. Kayser and A.S. Goldhaber, \prd {28}
(1983) 2341; B. Kayser, \prd {30}
(1984) 1023. \par
\bi{79}
W.S. Marciano and A. Sanda, \plb  {77} (1977) 303;
B.W. Lee and R. Shrock, \prd {60} (1977) 444;
S. Petcov, Sov. J. Nucl. Phys. {25} (1977) 340;
(E): Sov. J. Nucl. Phys. {25} (1977) 698;
K Fujikawa and R. Shrock, \prl {45} (1980) 963.\par
\bi{susynmm}
R. Barbieri, M.M. Guzzo, A. Masiero and D. Tommasini, \plb 252
(1990) 251.\par
\bi{88}
F. Gabbiani, A. Masiero and D.W. Sciama, \plb 259 (1991) 323.\par
\bi{cnd}
E. Roulet and D. Tommasini, \plb  256 (1991) 218;\hbup
D. Tommasini, \plb 297 (1992) 125.\par
\bi{enq-ma-rio}
K. Enqvist, A. Masiero and A. Riotto, \npb 373 (1992) 95.\par
\bi{masiero-riotto}
A. Masiero and A. Riotto, \plb 289 (1992) 73.\par
\bi{cnde6}
J. Maalampi and M. Roos, \plb  263 (1991) 437.\par
\interlinea
\bi{cosmlimlam}
B.A. Campbell, S. Davidson, J. Ellis and K. Olive,
\plb  256 (1991) 457; Astropart. Phys. 1 (1992) 77; \hbup
W. Fishler, G.F. Giudice, R.G. Leigh and S. Paban, \plb 258 (1991) 45;\hbup
H. Dreiner and G.G. Ross, \npb 410 (1993) 188.\par
\bi{dreiner-ross}
H. Dreiner and G.G. Ross, \npb 410 (1993) 188.\par
\bi{ibanez-quevedo}
L.E. Iba\~ nez and F. Quevedo, CERN-TH-6433/92.\par
\bi{susynovelties}
A. Masiero, DFPD 91/TH/11.\par
\bi{spontrbreak}
A. Masiero and J.W.F. Valle, \plb 251 (1990) 273.\par
\bi{msw-susy-roulet}
E. Roulet, \prd 44 (1991) R953.\par
\bi{msw-fc}
M.M. Guzzo, A. Masiero and S. Petcov, \plb 260 (1991)154;\hbup
V. Barger, R.J.N. Phillips and K. Wishnant, MAD/PH/648 (1991).\par
\bi{msw-e6-roulet}
E. Roulet, FERMILAB--PUB--91/206--A.\par
\bi{bfz}
S.M. Barr, E.M. Freire and A.Zee, \prl  65 (1990) 2626.\par
\bi{bf}
S.M. Barr and E.M. Freire, \prd  43 (1991) 2989.\par
\bi{babuetal}
K.S. Babu, D. Chang, W.-Y. Keung and I. Phillips, FERMILAB-PUB-92/56-T
(1992).\par
\bi{msw}
L. Wolfenstein, \prd 17 (1978) 2369;
S.P. Mikheyev and A.Y. Smirnov, Yad. Fiz. 42 (1985) 1441 [Sov. J. Nucl.
Phys. 42 (1985) 913].\par
\bi{smirnov93}
A.Y. Smirnov, preprint IC/93/359 (1993).\par
\bi{hotdm}
E.L. Wright \ea, Astr. J. 396 (1992) L13;\hbup
P. Davis, F.J. Summers and D. Schlegel, Nature 359 (1992) 393;\hbup
R.K. Schafer and Q. Shafi, preprint BA--92--28 (1992);\hbup
J.A. Holzman and J.R. Primack, Astr. J. 405 (1993) 428.\par
\bi{davis}
davis\par
\bi{17kev-msw}
L. Bento and J.W.F. Valle, \plb  264 (1991) 373;
J. Peltoniemi, A.Y. Smirnov and J.W.F. Valle, FTUV/92-6 (1992).\par
\bi{bento-valle}
L. Bento and J.W.F. Valle, \plb  264 (1991) 373.\par
\bi{pelt-smi-valle}
J. Peltoniemi, A.Y. Smirnov and J.W.F. Valle, FTUV/92-6 (1992).\par
\bi{pelt-tom-valle}
J. Peltoniemi, D. Tommasini and J.W.F. Valle, \plb 298 (1993) 383.\par

\biblitem{pdg}
Rev. of Part. Prop., \plb 239 (1990) 1. \par
\bi{fit}
E. Nardi, E. Roulet and D. Tommasini, SISSA 104/91/EP
                            (FERMILAB 91/207-A), submitted to \npb. \par

\bi{fit-fit6}
E. Nardi, E. Roulet and D. Tommasini, FERMILAB 91/207-A (1991),
FERMILAB 92/127-A (1992).\par

\insertbibliografia

\par\vfill\eject

\centerline{\elevenbf FIGURE CAPTION}

\bs\bs\bs

\no Figure 1. Contributions to the neutrino mass element $m_{\nu_i\nu_j}$
in the SSM with broken R--parity.

\bye